\documentclass[prl,preprint,amssymb,amsmath,nobibnotes,nofootinbib]{revtex4}
\usepackage{amsfonts}
\usepackage{revsymb}
\usepackage{graphicx,epsfig}
\usepackage{placeins}

\begin{document}
\title {Regularization of the second-order gravitational
 perturbations produced by a compact object}
\author{Eran Rosenthal}
\affiliation{
 Department of Physics, University of Guelph, Guelph, Ontario  N1G 2W1,
Canada}

\date{\today}

\begin{abstract} 
The equations for the second-order gravitational perturbations produced by a compact-object
have highly singular source terms at the point particle limit. 
At this limit the standard retarded solutions 
to these equations are ill-defined. 
Here we construct well-defined and physically meaningful
 solutions to these equations.
These solutions are important for practical calculations:
the planned gravitational-wave detector LISA requires 
preparation of waveform templates for the expected gravitational-waves. 
Construction of templates with desired accuracy for extreme mass ratio binaries,  
in which a compact-object inspirals towards a supermassive black-hole,  
requires calculation of the second-order gravitational 
perturbations produced by the compact-object. 
\end{abstract} 

\maketitle

In an extreme mass-ratio binary,  
where a compact-object (CO) of mass $\mu$ 
(e.g. a neutron star or a small black hole)
orbits a supermassive black hole of  mass $M$, 
the interaction between the CO and the gravitational perturbations that 
it produces exerts a force on the CO. 
This phenomenon of gravitational self-force (GSF) is responsible for  
the CO's gradual inspiral towards the supermassive black-hole. 
The small parameter $\mu/M$ allows treating the spacetime metric 
and the CO's orbit perturbatively. 
At the leading order of this approximation 
the orbit traces a geodesic 
in the background spacetime of 
the supermassive black-hole \cite{DEATH}. 
At the next order, the CO's interaction with 
the $O(\mu)$ first-order metric perturbations (FOMP) 
produces a first-order GSF that accelerates the CO
in the background spacetime \cite{QW,MST}. 
At the next order, the interaction with the $O(\mu^2)$ second-order 
metric perturbations (SOMP) produces a second-order GSF, and so on.

Extreme mass-ratio binaries (e.g., $M/\mu= 10^{5}$) 
are valuable sources of 
gravitational waves (GW) that could be detected 
by the planned Laser Interferometer Space Antenna (LISA) \cite{LISA}.
Detection of these sources and determination of their parameters  
using matched-filtering data-analysis  
techniques requires preparation of gravitational  
waveform templates for the expected GW. 
Here one of the main challenges is
 to calculate the GW's accumulating 
phase, e.g. a wave-train from one year of inspiral can contain  
 about $10^5$ GW cycles \cite{TC}. 
Successful determination of the binary parameters 
using matched-filtering techniques 
often requires GW templates in which the phase error   
is less than one radian over a year \cite{BC}. 

The following simple scaling argument reveals how 
 the phase scales with $\mu$, and allows us to 
quantify the desired degree of accuracy
for the metric perturbation expansion (see also \cite{BURKO,DETWEILER}). 
For simplicity consider a CO that gradually inspirals 
 between two otherwise circular geodesic orbits 
in a strong field region of a Schwarzschild black hole.
We are interested in estimating the effect of the GSF 
on the accumulated phase of the emitted GW.
Due to the GSF the orbital frequency slowly changes from 
its value at an initial time, and 
after time $t$ has elapsed
this frequency has shifted by approximately $\dot{\omega}t$ from its initial value, where
 $\dot{\omega}\equiv\frac{d\omega}{dt}$, and $t$ denotes the elapsed time (from initial time) 
in Schwarzschild coordinates.
We denote $\Delta\phi$ the part of the phase shift of the GW (between two fixed times)
which is induced by the shift in the orbital frequency. 
Recalling that the GW frequency is proportional to the orbital frequency we find that  
after an inspiral time $\Delta t_{ins}$, the phase shift 
$\Delta\phi$ is approximately proportional to $\Delta t^2_{ins} \dot{\omega}$.
We shall now find how the quantities in this expression scale with $\mu$.
Let us consider first $\dot{\omega}$. We write this quantity as 
$\dot{\omega}=\frac{d\omega}{dE}\dot{E}$, where $E$ denotes 
the particle's energy per unit mass. 
Notice that $\frac{d\omega}{dE}$ is independent of $\mu$ 
(it is obtained from an analysis of a circular geodesic worldline) while 
$\dot{E}$ is determined from the four-acceleration and therefore depends on $\mu$. 
The first-order GSF produces the first-order terms in the expansions of 
$\dot{E}$ and $\dot{\omega}$; these terms are 
denoted here $\dot{E}_1$ and $\dot{\omega}_1$, respectively. Since $\dot{E}_1$ 
 scales like $\mu$ we find that $\dot{\omega}_1$ 
is $O(\mu M^{-3})$.
At the leading order the inspiral is driven by 
the first-order GSF and therefore the inspiral time 
 $\Delta t_{ins}$ is of the order $\Delta{E}\dot{E}^{-1}=O(M^2 \mu^{-1})$,
where $\Delta{E}$ is the energy difference between initial and final orbits.
After  $\Delta t_{ins}$  the term  $\dot{\omega}_1$ will give rise to 
a phase shift of order $\Delta t^2_{ins} \dot{\omega}_1=O(M/\mu)$.
The second-order GSF produces second-order terms in the expansions of $\dot{E}$, and 
 $\dot{\omega}$. These terms are 
denoted here $\dot{E}_2$ and $\dot{\omega}_2$, respectively.
Since  $\dot{E}_2$ scales like $\mu^2$ it gives rise to an $\dot{\omega}_2$ 
which is $O(\mu^2 M^{-4})$. 
After $\Delta t_{ins}$ the term $\dot{\omega}_2$ will produce  
a phase shift of order $\Delta t^2_{ins} \dot{\omega}_2=O[(M/\mu)^0]$.
Therefore, a calculation of $\Delta\phi$ to the desired 
accuracy of order $(M/\mu)^0$ (needed for LISA data analysis)
requires the calculation of the CO's interaction with its own SOMP.

The challenge of constructing long waveform templates with 
 $O[(M/\mu)^0]$ accuracy provides a practical motivation for the study
of SOMP in this article. Furthermore, 
construction of SOMP will extend the applicability of the perturbative scheme 
to binaries with smaller $M/\mu$ mass-ratios. 

Consider the limit where the spatial dimensions
 of the CO approach zero;
we shall refer to this limit as the point particle limit
(Below we shall make more precise definitions of the 
CO and the limiting process that we use.).
In this limit the SOMP away from the CO satisfy the following 
equation
\begin{equation}\label{2ndgrav} 
D[l]= \nabla h\nabla h \ \&\  h\nabla\nabla h\,.  
\end{equation} 
Here we used a schematic notation, where 
the SOMP are denoted $\mu^2 l$, 
$D$ denotes a linear partial differential operator, 
the retarded FOMP are denoted $\mu h$,
 $\nabla$ denotes a covariant derivative with respect 
 to the background metric, and $\&$ denotes ``and terms of the form ...''. 
In the point particle limit 
 (certain components of) the FOMP in the Lorenz gauge
diverge as $r^{-1}$, 
where $r$ denotes the spatial distance from the object.  
Here we encounter a problem, since the $O(r^{-1})$ divergent 
 behavior of $h$  
implies that the source term of Eq. (\ref{2ndgrav}) diverges as $r^{-4}$. 
A naive attempt to construct the standard retarded solution to Eq. (\ref{2ndgrav})
 by imposing Lorenz gauge conditions on $l$, 
and then integrating the resultant wave equation  
using the retarded Green's function, results in an integral that diverges 
 at every point in spacetime. A similar problem in a scalar 
toy-model has recently been studied \cite{scalar}.
In this article we develop a regularization method
for the construction of well-defined and physically meaningful solutions to
 Eq. (\ref{2ndgrav}).

We should mention here another regularization problem
coming from infinity.   
For an infinitely long 
worldline, the leading asymptotic behavior of the source term in 
Eq. (\ref{2ndgrav}) is $O(r^{-2})$. 
This behavior at infinity renders the retarded solution ${l}$
divergent \footnote{Asymptoticly 
$h$ has a form of a gravitational wave. Therefore
the dominant terms in an asymptotic expansion of $\nabla h$ 
decay like $r^{-1}$. This implies that  the source term in 
Eq. (\ref{2ndgrav}) is asymptotically $O(r^{-2})$. 
Notice that this source term has a static  $O(r^{-2})$ 
component which does not vanish after averaging over time.
An attempt to construct the retarded solution to 
Eq. (\ref{2ndgrav}) with this static term present 
yields a divergent expression (even if we resolve the difficulty with the singularity 
in the vicinity of the worldline).  
Ori has recently suggested 
a resolution to this problem \cite{ORI}.}. 
This issue, however, lies outside the scope of this article. Here   
we assume that a regularization at infinity had been carried out.

For simplicity we choose the CO to be  
 a Schwarzschild black hole, which moves 
in a vacuum background geometry characterized by
 length scales that are much larger than $\mu$. Designating the length scales 
associated with the Riemann curvature
 tensor of the background geometry with $\{{\cal R}_i\}$, and denoting 
 ${\cal R}=\min\{{\cal R}_i\}$, we express the above restriction 
as $\mu\ll\cal{R}$.

Our analysis is based
on the method of matched asymptotic expansions (see e.g. \cite{TH,PR}),
in which different approximate solutions to Einstein's field equations are  
obtained in different, overlapping, regions of spacetime. These solutions
are then matched in their common region of validity.
Consider the following decomposition of spacetime into
an internal zone that lies within a worldtube 
which surrounds the black hole and extends out to $r=r_I(\cal{R})$
 such that $r_I\ll\cal{R}$, and an external zone that lies 
 outside another worldtube at $r=r_E(\mu)$, such that $\mu\ll r_E$. 
The interior of this inner worldtube is denoted $S$.
By virtue of $\mu\ll\cal{R}$ we choose $r_E$ to be much smaller than $r_I$ such 
that there is an overlap between the above mentioned regions in $r_E<r<r_I$.
We refer to this overlap region as the buffer zone 
(in the buffer zone $r$ can be of order $\sqrt{\mu \cal{R}}$). 

In the external zone we decompose the full spacetime metric $g^{full}_{\mu\nu}$
into a background metric $g_{\mu\nu}$ (e.g. a spacetime of a supermassive black hole),
 and a sequence of perturbations that are produced by the  Schwarzschild
 black hole (with mass $\mu$), reading
\begin{equation}\label{metricdec}
g^{full}_{\mu\nu}(x)=g_{\mu\nu}(x)+\mu h_{\mu\nu}(x)+\mu^2 l_{\mu\nu}(x)+O(\mu^3)\,.
\end{equation}
Here the dependence on $\mu$ is only through the explicit powers $\mu^i$.
Throughout we use the background metric $g_{\mu\nu}$ to 
raise and lower tensor indices, and to evaluate covariant derivatives.
Substituting expansion (\ref{metricdec}) into Einstein's field equations in vacuum
yields linear perturbation equations for $h_{\mu\nu}$ and $l_{\mu\nu}$.
These equations are valid for $x\not\in S$. 
Here we shall be interested in the limit $\mu\rightarrow 0$.
Note that this limit has no effect on the forms 
of the perturbations equations. Yet their region of validity 
is affected since we let $r_E(\mu)\rightarrow 0$ as $\mu \rightarrow 0$.
In this limit the perturbation equations read 
\begin{eqnarray}\label{h1eq}
&&D_{\mu\nu}[\bar{h}]=0\ ,\ x\not\in z(\tau)\,, \\\label{h2eq}
&&D_{\mu\nu}[\bar{l}]=S_{\mu\nu}[\bar{h}]\ ,\ x\not\in z(\tau)\,.
\end{eqnarray}
Here $z(\tau)$ is a timelike worldline, 
 where $\tau$ denotes proper time with respect to the background metric.
An overbar denotes the trace-reversal operator 
$\bar{h}_{\mu\nu}=h_{\mu\nu}-(1/2)g_{\mu\nu}h_\alpha^{\ \,\alpha}$;
for brevity we have omitted tensorial indices inside the squared brackets. 
$D_{\mu\nu}$ and $S_{\mu\nu}$ are obtained from an expansion of the
 full Ricci tensor. By substituting $g_{\mu\nu}+\delta g_{\mu\nu}$
into the Ricci tensor we obtain 
$R^{full}_{\mu\nu}=R_{\mu\nu}^{(0)}+R_{\mu\nu}^{(1)}[\delta g]+
R_{\mu\nu}^{(2)}[\delta g]+O(\delta g^3)$. (For explicit expressions 
for the terms in this expansion see e.g. \cite{MTW}. 
We adopt the sign convention of this reference for the Riemann tensor). 
To simplify the notation we denote $\bar{R}_{\mu\nu}^{(1)}[h]$ by $D_{\mu\nu}[\bar{h}]$, 
and denote $-\bar{R}_{\mu\nu}^{(2)}[h]$ by $S_{\mu\nu}[\bar{h}]$, where
$h_{\mu\nu}$ is expressed in terms of $\bar{h}_{\mu\nu}$.

By imposing the Lorenz gauge conditions
$\bar{h}^{\mu\nu}_{\ \ \, ;\nu}=0$, and matching with the internal-zone solution,  
D'Eath showed \cite{DEATH} that $h_{\mu\nu}$  
is identical to the retarded FOMP  
produced by a unit-mass point particle tracing  
the worldline $z(\tau)$. At this leading order of approximation $z(\tau)$
 is a geodesic of the background 
spacetime \cite{DEATH}, denoted $z_G(\tau)$.
Substituting $z(\tau)=z_G(\tau)$ in Eq. (\ref{h1eq}) provides us 
with sufficient accuracy for the calculation
of the FOMP. However, to calculate the SOMP we must account for the $O(\mu)$ 
acceleration of $z(\tau)$ due to the first-order GSF corrections.
To account for these corrections we 
use the gauge dependence of the first-order GSF \cite{BO},
 and choose a convenient gauge in which the first-order GSF vanishes. 
In this gauge, the geodesic worldline $z_G(\tau)$ is  
 sufficiently accurate for the construction of the SOMP. 

To spell out the desired gauge conditions we 
examine the expression for the $O(\mu)$ acceleration which is 
produced by the first-order GSF in the Lorenz gauge \cite{DW}
\begin{equation}\label{selfaccel}
a^{\mu}=-\frac{\mu}{2}(g^{\mu\nu}+u^{\mu}u^{\nu})u^{\rho}u^{\eta}
(2\nabla_\rho h^{R}_{\eta\nu}-\nabla_{\nu}h^{R}_{\rho\eta})\,.
\end{equation}
Here all quantities are evaluated on the worldline.
We use the Detweiler-Whiting decomposition ${h}_{\mu\nu}={h}^{S}_{\mu\nu}+{h}^{R}_{\mu\nu}$,
where $h^{R}_{\mu\nu}$ and $h^{S}_{\mu\nu}$ are certain regular 
and singular potentials, respectively 
(for their definitions and properties see \cite{DW}).
Consider now a regular gauge transformation generated by 
a vector field $\xi^\mu$. In 
the new gauge the FOMP, $h^F_{\mu\nu}$, are given by  
$h^F_{\mu\nu}=h^{S}_{\mu\nu}+h^{R(F)}_{\mu\nu}$, where
$h^{R(F)}_{\mu\nu}\equiv h^{R}_{\mu\nu}+\xi_{\mu;\nu}+\xi_{\nu;\mu}$
(The gauge invariance of $h^{S}_{\mu\nu}$ follows from the 
analysis in Ref. \cite{BO}.).

Many gauge choices can provide us with the desired requirement $a^{\mu}=0$.
For simplicity we impose the following gauge conditions on the worldline:
\begin{equation}\label{fermigauge}
\left[h^{R(F)}_{\mu\nu}\right]_{z_G(\tau)}=0\ ,\ 
\left[\nabla_\rho h^{R(F)}_{\mu\nu}\right]_{z_G(\tau)}=0\,.
\end{equation}
We shall refer to this gauge as the Fermi gauge. We may now replace 
the restrictions $x\not\in z(\tau)$ with 
$x\not\in z_G(\tau)$ in Eqs. (\ref{h1eq},\ref{h2eq}).
Notice that the gauge conditions (\ref{fermigauge}) are specified only on $z_G(\tau)$.  
To complete the gauge construction we 
adopt an arbitrary continuation of $\xi_\mu$ to the entire spacetime 
(e.g., continuation along future null-cones based on $z_G(\tau)$ 
in a manner that preserves causality \footnote{
To construct this continuation consider the
 future null-cones $\Sigma_\tau$ emanating from the worldline $z_G(\tau)$. 
Here we focus on a local neighborhood of the worldline in which
these null-cones do not intersect each other.
For an arbitrary point $z_G({\tau^-})$ 
one may choose an arbitrary continuation of $\xi^\mu$ on $\Sigma_{\tau^-}$,
such that $\xi^\mu$ decays to zero away from the worldline.
In this way the constructed gauge preserve causality in the following sense:
The perturbations on the null-cones $\Sigma_{\tau}$ for $\tau\le \tau^-$ will
 remain unchanged if one modifies the worldline for 
$\tau>\tau^-$. Such a modification of the worldline 
is possible by introducing additional GW that 
interact with the worldline for $\tau>\tau^-$.}).

Before solving  
Eq. (\ref{h2eq}) we study the singular properties of $S_{\mu\nu}$ near
$z_G(\tau)$. Consider expanding $S_{\mu\nu}$  
 in the vicinity of $z_G(\tau)$, on a family of hypersurfaces $\tau=const$ 
that are generated by geodesics that are normal to 
the worldline. We use 
Fermi normal coordinates based on $z_G(\tau)$, thus obtaining simple expressions. 
We find that
\begin{equation}\label{sourcexpan}
S_{\mu\nu}(x)\stackrel{*}=A_{\mu\nu}r^{-4}+O(r^{-2})\,.
\end{equation}
Here $\stackrel{*}=$ denotes equality in Fermi normal coordinates.
We have 
 $A_{\mu\nu}\equiv 4u_{\mu}u_{\nu}+7\eta_{\mu\nu}-14\Omega_{\mu}\Omega_{\nu}$, 
$\eta_{\mu\nu}$ denotes the Minkowski metric, 
$r=\sqrt{\delta_{ab}x^a x^b}$, where $x^a$ denote the 
spatial Fermi coordinates ($a,b$ take the values $\{1,2,3\}$);
 $u^\mu\stackrel{*}=\delta^\mu_0$ is a vector
field, $\Omega^a\stackrel{*}=x^a/r$,  $\Omega^0\stackrel{*}=0$, 
and $\Omega_\nu\stackrel{*}=g_{\nu\mu}\Omega^\mu$.
The absence of $O(r^{-3})$ terms in Eq. (\ref{sourcexpan}) follows
from the gauge conditions (\ref{fermigauge}) together with the fact 
that $z_G(\tau)$ is a geodesic worldline.

To tackle the $r^{-4}$ singularity of $S_{\mu\nu}$
we decompose $\bar{l}_{\mu\nu}$ into two tensor potentials, reading 
\begin{equation}\label{psol}
\bar{l}_{\mu\nu}=\bar{\psi}_{\mu\nu}+\delta \bar{l}_{\mu\nu}\,,
\end{equation}
where $\bar{\psi}_{\mu\nu}$ satisfies $D_{\mu\nu}[\bar{\psi}]
\stackrel{*}=A_{\mu\nu}r^{-4}+O(r^{-2})$.
Notice that this equation has 
the same $r^{-4}$ singular source term as Eq. (\ref{h2eq}).
Here, however, we do not impose restrictions on the terms of order $O(r^{-2})$.
Suppose that we can construct a solution $\bar{\psi}_{\mu\nu}$, then by subtracting
$D_{\mu\nu}[\bar{\psi}]$ from both sides of Eq. (\ref{h2eq}) we obtain 
\begin{equation}\label{deltaleq}
D_{\mu\nu}[\delta \bar{l}]=\delta S_{\mu\nu}\ \  ,x\not\in z_G(\tau)\,.
\end{equation}
Here $\delta S_{\mu\nu}\equiv S_{\mu\nu}-D_{\mu\nu}[\bar{\psi}]$. By construction 
$\delta S_{\mu\nu}$ diverges as $r^{-2}$, while $S_{\mu\nu}$ in 
Eq. (\ref{h2eq}) diverges as $r^{-4}$. In this 
sense Eq. (\ref{deltaleq}) is simpler than Eq. (\ref{h2eq}).

To construct $\bar{\psi}_{\mu\nu}$  we use a linear combination of 
terms that are quadratic in $\bar{h}^F_{\mu\nu}$.
Since $\bar{h}^F_{\mu\nu}$ diverges as $r^{-1}$ we find 
that by applying the operator $D_{\mu\nu}$ to terms
which are quadratic in $\bar{h}^F_{\mu\nu}$ we obtain 
terms which diverge as $r^{-4}$. First we construct four independent
quadratic tensor fields:
$\varphi^{A}_{\mu\nu}=\bar{h}^{F\rho}_{\ \ \ \mu}
\bar{h}^{F}_{\rho\nu}$, $\varphi^{B}_{\mu\nu}=\bar{h}^{F\rho}_{\ \ \ \rho}
\bar{h}^{F\ \ \ \ }_{\mu\nu}$, $\varphi^{C}_{\mu\nu}=(\bar{h}^{F\eta\rho}
\bar{h}^{F}_{\ \ \eta\rho})g_{\mu\nu}$,  
$\varphi^{D}_{\mu\nu}=\left(\bar{h}^{F\rho}_{\ \ \ \rho}\right)^2
g_{\mu\nu}$. Combining these terms we may 
construct the desired $\bar{\psi}_{\mu\nu}$, which reads
\begin{equation}\label{psiexplicit}
\bar{\psi}_{\mu\nu}=\frac{1}{64}\left[2(c_A\varphi^{A}_{\mu\nu}+
c_B\varphi^{B}_{\mu\nu})-7(c_C\varphi^{C}_{\mu\nu}+
c_D\varphi^{D}_{\mu\nu})\right]\,.
\end{equation}
Here  the constants $c_A,c_B,c_C,c_D$ must satisfy
$c_A+c_B=1$, $c_C+c_D=1$, 
but are otherwise arbitrary. 
Roughly speaking, $\bar{\psi}_{\mu\nu}$ 
captures the asymptotic behavior of the Schwarzschild 
solution at second order. Due to this property $\bar{\psi}_{\mu\nu}$
 is an {\em exact solution} to Eq. (\ref{h2eq}) 
for the case of a flat background spacetime.

Having constructed $\bar{\psi}_{\mu\nu}$,  
we now face the problem of solving Eq. (\ref{deltaleq}).
For this purpose we invoke a purely second-order 
gauge transformation of the form 
$x^\mu\rightarrow x^{\mu}-\mu^2 \xi^{\mu}_{(2)}$
that allows us to impose the Lorenz gauge conditions 
$\delta\bar{l}^{\mu\nu}_{\ \ \ ;\nu}=0$.
First we seek a particular retarded solution to Eq. (\ref{deltaleq}). 
Here it is useful to remove the restriction $x\not\in z_G(\tau)$,
and continue $\delta S_{\mu\nu}$ to the worldline $z_G(\tau)$.
Clearly a solution to the continued equation also satisfies the original 
equation (i.e. with the worldline excluded). 
We choose the simplest continuation by demanding
that no additional singularities (e.g. delta functions) 
 are introduced on the worldline i.e., the only singularities 
on the worldline must be those originating from a local 
expansion of $S_{\mu\nu}$. 
Eq. (\ref{deltaleq}) now reads 
\begin{equation}\label{deltah2final}
\Box \delta\bar{l}_{\mu\nu}+2R^{\eta\ \rho}_{\ \mu\ \nu}
\delta\bar{l}_{\eta\rho}=-2\delta S_{\mu\nu} \,.
\end{equation}
Here $\Box\equiv g^{\alpha\beta}\nabla_{\alpha}\nabla_{\beta}$.
We define $\delta\bar{l}_{\mu\nu}$ to be the retarded solution 
to Eq. (\ref{deltah2final}). 
Since $\delta S_{\mu\nu}$ diverges only as $r^{-2}$ 
its retarded solution has a finite contribution originating
from the vicinity of $z_G(\tau)$.

By definition the retarded solution $\delta\bar{l}_{\mu\nu}$  
satisfies Eq. (\ref{deltah2final}). However, we 
still have to verify that it also satisfies Eq. (\ref{deltaleq}). 
This equation will be satisfied if $\delta\bar{l}_{\mu\nu}$ satisfies the Lorenz
gauge conditions.
To investigate this point one can apply the divergence operator to Eq. (\ref{deltah2final}), 
thereby constructing a differential equation for $\nabla^\nu\delta\bar{l}_{\mu\nu}$.
The fact that $\nabla^\nu\delta\bar{l}_{\mu\nu}=0$ then follows from 
the properties of the source term of this equation, namely $-2\nabla^\nu\delta S_{\mu\nu}$. 
Using a perturbation expansion of the Bianchi identities one finds 
that $\nabla^\nu\delta S_{\mu\nu}\equiv 0$ for $x\not\in z_G(\tau)$. 
This property together with an analysis of $\nabla^\nu\delta S_{\mu\nu}$  
as one approaches the worldline reveals that   
the retarded solution to Eq. (\ref{deltah2final}) 
 satisfies the Lorenz gauge conditions,
and therefore it also satisfies Eq. (\ref{deltaleq})
as required. (The full analysis will be given elsewhere \cite{ER}.) 
  
So far we have constructed 
a particular solution (\ref{psol}) to Eq. (\ref{h2eq}).
We now construct the general solution $\bar{l}^{G}_{\mu\nu}$ to this equation. 
Later we shall impose a set of additional requirements 
on $\bar{l}^{G}_{\mu\nu}$, and thereby obtain a particular solution
which is physically meaningful.
Since Eq. (\ref{h2eq}) is valid for $x\not\in z_G(\tau)$,
we find that we can construct a new solution by adding to $\bar{l}_{\mu\nu}$
a potential that satisfies a {\em semi-homogeneous} equation ,i.e.,
a homogeneous equation for $x\not\in z_G(\tau)$, reading
\begin{equation}\label{sheq} 
D_{\mu\nu}[\bar{l}^{SH}]=0 \,\,\, ,x\not\in z_G(\tau)\,. 
\end{equation} 
The general solution to Eq. (\ref{h2eq}) is given by $\bar{l}^{G}_{\mu\nu}\equiv 
\bar{l}^{SH}_{\mu\nu}+\bar{l}^{}_{\mu\nu}$, where
$\bar{l}^{SH}_{\mu\nu}$ is the general solution to Eq. (\ref{sheq}). 
 
The set of additional requirements to be imposed on $\bar{l}^{G}_{\mu\nu}$
can be expressed as requirements on $\bar{l}^{SH}_{\mu\nu}$.
Denoting $\bar{\gamma}_{\mu\nu}$ the semi-homogeneous potential that 
satisfies these requirements, 
we express the desired physical solution $\bar{l}^{P}_{\mu\nu}$ as  
\begin{equation}\label{lp}  
\bar{l}^{P}_{\mu\nu}=\bar{l}^{}_{\mu\nu}+\bar{\gamma}_{\mu\nu}=
 \bar{\psi}_{\mu\nu}+\delta\bar{l}^{}_{\mu\nu}+\bar{\gamma}_{\mu\nu}\,.
\end{equation}
We consider the following additional requirements: 
 (i) gauge conditions, (ii) causality requirements, 
(iii) global boundary conditions, and (iv) 
boundary conditions as $x\rightarrow z_G(\tau)$.
(i) We impose Lorenz gauge conditions on $\bar{\gamma}_{\mu\nu}$.
    Eq. (\ref{sheq}) now takes the form
\begin{equation}\label{gammaeq}
\Box \bar{\gamma}_{\mu\nu}+2R^{\eta\ \rho}_{\ \mu\ \nu}\bar{\gamma}_{\eta\rho}=0
 \,\,\,  ,x\not\in z_G(\tau)\,.
\end{equation}
(ii) We define $\bar{\gamma}_{\mu\nu}$ 
to be a retarded solution to Eq. (\ref{gammaeq}). Recall that 
$\delta\bar{l}^{}_{\mu\nu}$ is the retarded solution to 
Eq. (\ref{deltah2final}), and $\bar{\psi}_{\mu\nu}$
is completely determined by $\bar{h}^{F}_{\mu\nu}$.  
This construction implies 
that if we prescribe initial data for the metric perturbations 
on a spacelike hypersurface $\Sigma_0$,
then the physical solution $\bar{l}^{P}_{\mu\nu}(x)$ at the future of  
$\Sigma_0$ is unaffected by an arbitrary modification of the 
initial data outside $J^-(x)\cap\Sigma_0$, and in this sense 
the second-order solution $\bar{l}^{P}_{\mu\nu}(x)$ is 
manifestly causal.
(iii) We demand that the only source of $\bar{\gamma}_{\mu\nu}$ is the black-hole itself, 
thus excluding any incoming waves. 
(iv) Let us consider once more a small 
but finite mass $\mu$. Here, 
Eq. (\ref{gammaeq}) is valid at $x\not\in S$. 
Following D'Eath's analysis of FOMP \cite{DEATH} we 
express a solution to Eq. (\ref{gammaeq}) using a Kirchhoff representation. 
 Denoting $\Sigma_E$ the boundary of $S$, 
we express $\bar{\gamma}_{\mu\nu}$ in terms of the following surface integral
\begin{eqnarray}\label{kirchhoff}
\bar{\gamma}_{\mu\nu}(x)&=&-\frac{1}{4\pi}\int_{\Sigma_E(\mu)}\Big(
G^{ret}_{\mu\nu\alpha'\beta'}[x|x']
\nabla^{\epsilon'}\bar{\gamma}^{\alpha'\beta'}(x')\\\nonumber
&&-\bar{\gamma}^{\alpha'\beta'}(x')\nabla^{\epsilon'}G^{ret}_{\mu\nu\alpha'\beta'}[x|x']\Big)d\Sigma_{\epsilon'}\,.
\end{eqnarray}
Here $G^{ret}_{\mu\nu\alpha'\beta'}[x|x']$
denotes the retarded Green's function 
(for its definition and properties see e.g. \cite{PR}),
$d\Sigma_{\epsilon'}$ denotes an outward directed three-surface element on 
$\Sigma_E$. Consider substituting an expansion (in powers of $r$) of 
a given potential $\bar{\gamma}^{\alpha\beta}$
 into Eq. (\ref{kirchhoff}), and then taking the limit $\mu\rightarrow 0$. 
Recall that in this limit we have $r_E(\mu)\rightarrow 0$, and 
notice that $d\Sigma_\epsilon$ scales as $r_E^2$. Therefore,  
only the diverging terms (as $r\rightarrow 0$) in the expansion
can contribute to $\bar{\gamma}_{\mu\nu}(x)$ 
in the limit. These {\em divergent boundary conditions} 
for Eq. (\ref{gammaeq}) are obtained 
from Eq. (\ref{lp}) together with an analysis 
of the divergent behavior of $\bar{\psi}_{\mu\nu}$, $\delta\bar{l}_{\mu\nu}$, 
and $\bar{l}^{P}_{\mu\nu}$ in near $r=0$. 

First we expand $\bar{\psi}_{\mu\nu}$ near $r=0$ and obtain
$\bar{\psi}_{\mu\nu}\stackrel{*}=-\frac{1}{4r^2}[2u_{\mu}u_{\nu}+7\eta_{\mu\nu}]+O(r^0)$.
Next we consider $\delta\bar{l}_{\mu\nu}$. 
Solving Eq. (\ref{deltah2final}) approximately in the vicinity of $z_G(\tau)$   
reveals that $\delta\bar{l}_{\mu\nu}$ is bounded as $r\rightarrow 0$ \cite{ER}. 
Finally, we expand $\bar{l}^{P}_{\mu\nu}$ using   
the internal-zone solution, which is discussed next. 

We now calculate the desired diverging 
 terms in the expansion of $\bar{l}^{P}_{\mu\nu}$.
For this purpose we use the method of 
matched asymptotic expansions.
For simplicity we assume that all 
the length scales characterizing
the background spacetime $\{{\cal R}_i\}$ are of the same order of magnitude 
$\cal{R}$.  
Expanding the full spacetime metric in the internal-zone using 
the smallness of $r/{\cal R}$ and $\mu/\cal{R}$ yields
\begin{equation}\label{intexp}
g^{full}_{\mu\nu}=g^{Sch}_{\mu\nu}
+{\cal R}^{-1}g^{(1)}_{\mu\nu}+{\cal R}^{-2}g^{(2)}_{\mu\nu}
+O({\cal R}^{-3})\,.
\end{equation}
Here $g^{Sch}_{\mu\nu}$ is the metric of the Schwarzschild black-hole.
Recall that in the buffer-zone both Eq. (\ref{intexp}) and Eq. (\ref{metricdec}) 
are valid, and should therefore match with each other. 
To match the internal-zone metric with the external-zone metric
we further expand Eqs. (\ref{metricdec},\ref{intexp}) in the buffer-one. 

First we consider the internal-zone expansion.
The first term in
Eq. (\ref{intexp}) is the Schwarzschild metric  $g^{Sch}_{\mu\nu}$.
In isotropic cartesian coordinates this metric takes the form 
 $ds^2=-(2\rho-\mu)^{2}(2\rho+\mu)^{-2}dt^2+
[1+\mu(2\rho)^{-1}]^4(dx^2+dy^2+dz^2)$, where $\rho^2=x^2+y^2+z^2$. 
In the buffer-zone $\mu\ll \rho$ and therefore 
the Schwarzschild metric can be approximated with an asymptotic expansion in
powers of $\mu \rho^{-1}$ giving
\begin{equation}\label{schasym}
g^{Sch}_{\mu\nu}=\eta_{\mu\nu}+
\mu{\rho}^{-1}H^{1}_{\mu\nu}+
\mu^2\rho^{-2}H^{2}_{\mu\nu}+O(\mu^3\rho^{-3})\,,
\end{equation}
where $H^{1}_{\mu\nu}=2(\eta_{\mu\nu}+2\delta_\mu^0\delta_\nu^0)$,
$H^{2}_{\mu\nu}=1/2(3\eta_{\mu\nu}-\delta_\mu^0\delta_\nu^0)$.
The next term in Eq. (\ref{intexp}) is ${\cal R}^{-1}g^{(1)}_{\mu\nu}$.
In a suitable gauge this term vanishes identically \cite{TH}. 

Next we consider the external-zone expansion (\ref{metricdec}). 
Recall that in the buffer-zone $r\ll{\cal R}$.
We therefore expand the terms in Eq. (\ref{metricdec})
in powers of $r{\cal{R}}^{-1}$.
The first term in Eq. (\ref{metricdec}) is the background metric $g_{\mu\nu}$. 
Expressing this metric with Fermi normal coordinates based on the worldline, and expanding it 
in powers of $r{\cal{R}}^{-1}$ gives 
$g_{\mu\nu}\stackrel{*}=\eta_{\mu\nu}+O({\cal R}^{-2} r^2)$.
Expanding the next term in (\ref{metricdec}) gives 
$\mu h_{\mu\nu}^F\stackrel{*}={2}\mu{r}^{-1}(\eta_{\mu\nu}+2 u_\mu u_\nu)+O(\mu{\cal R}^{-2}r^1)$,
 and expanding the third term $\mu^2{l}^{P}_{\mu\nu}$ gives a sum of a schematic form $\mu^2\sum_{i=0}^\infty{\cal{R}}^{-i}r^{i-2}$.
Recall that here we are only concerned with divergent boundary conditions that are encoded in 
the $i=0,1$ terms. 

First let us consider matching the first three terms 
 in the expansion of the Schwarzschild metric (\ref{schasym}), all these terms 
 scale like ${\cal R}^0$. 
From the above expansions we find that both  
$g^{Sch}_{\mu\nu}$ and $g_{\mu\nu}$ have 
the same leading term -- $\eta_{\mu\nu}$. 
The next term in Eq. (\ref{schasym}) is $\mu{\rho}^{-1}H^{1}_{\mu\nu}$ this term 
coincides with the leading order expansion of $\mu h_{\mu\nu}^F$,
 where we identify $\rho$ with $r$. 
The next term in Eq. (\ref{schasym}) provide us 
with the desired $i=0$ term in the expansion of $\mu^2{l}^{P}_{\mu\nu}$.
We find that this term is identical to the leading term in the expansion of $\psi_{\mu\nu}$.

Next let us consider matching the terms that scale like ${\cal{R}}^{-1}$.
Since ${\cal R}^{-1}g^{(1)}_{\mu\nu}=0$ all the terms that scale 
like ${\cal R}^{-1}$ vanish. This conforms with the fact 
 that the expansions of $g_{\mu\nu}$ and $\mu h_{\mu\nu}^F$ do not contain terms
that scale like ${\cal R}^{-1}$. This also 
implies that in the expansion 
of $\mu^2{l}^{P}_{\mu\nu}$ 
the term with the schematic form $\mu^2{\cal{R}}^{-1}r^{-1}$ is zero.
Combining the $i=0$ and $i=1$ terms in the expansion of $\mu^2{l}^{P}_{\mu\nu}$ 
we conclude that
$\bar{l}^{P}_{\mu\nu}\stackrel{*}=-\frac{1}{4r^2}[2u_{\mu}u_{\nu}+7\eta_{\mu\nu}]+O(r^0)$.

Combining the expansions 
of $\bar{\psi}_{\mu\nu}$ and $\bar{l}^{P}_{\mu\nu}$ near $r=0$, together 
with the fact that $\delta\bar{l}_{\mu\nu}$ is bounded in the vicinity of the worldline,
and using Eq. (\ref{lp}) we find that
$\bar{\gamma}_{\mu\nu}$ does not diverge as $x\rightarrow z_G(\tau)$. 
As was previously discussed only divergent boundary conditions 
 (as $r\rightarrow 0$) can produce non-vanishing
contributions to $\bar{\gamma}_{\mu\nu}(x)$. 
Since we found that $\bar{\gamma}_{\mu\nu}$ does 
not diverge in the vicinity of the worldline we conclude that 
$\bar{\gamma}_{\mu\nu}\equiv 0$.
By virtue of Eq. (\ref{lp}) we finally conclude that the physical SOMP are
given by
\begin{equation}\label{final}
\bar{l}^{P}_{\mu\nu}=\bar{\psi}_{\mu\nu}+\delta\bar{l}_{\mu\nu}\,.
\end{equation}
Here $\bar{\psi}_{\mu\nu}$ is given by Eq. (\ref{psiexplicit}), 
and $\delta\bar{l}_{\mu\nu}$ is the retarded solution to 
Eq. (\ref{deltah2final}). By construction $\bar{l}^{P}_{\mu\nu}$ 
satisfies the previously mentioned requirements. 
In particular it is manifestly causal, 
and it matches with the internal-zone solution.  
Moreover, Eq. (\ref{final}) provides a simple
covariant prescription for the construction of the SOMP, without
any reference to a particular (background) coordinate system. 
This allows a considerable amount 
of flexibility in the construction of the SOMP.  

\section*{Acknowledgments}

I am grateful to Amos Ori and Eric Poisson for numerous valuable discussions.
This work was supported in part by the Natural Sciences and Engineering
Research Council of Canada, and also in part by 
The Israel Science Foundation (grant no. 74/02-11.1).

 
\end{document}